\def\year{2017}\relax
\documentclass[letterpaper]{article} 
\usepackage{aaai17}  
\usepackage{times}  
\usepackage{helvet}  
\usepackage{courier}  
\usepackage{url}  
\usepackage{graphicx}  
\usepackage{color}
\usepackage{nameref}
\usepackage{amsmath}
\usepackage{float}
\begin{document}
%
\title{A Data-driven Approach Towards Human-robot\\
Collaborative Problem Solving in a Shared Space}

\author{Michael Wollowski  \and Carlotta Berry \and Ryder Winck  \and Alan Jern  \and\\
{\bf \Large David Voltmer  \and Alan Chiu \and Yosi Shibberu}\\
Rose-Hulman Institute of Technology\\
5500 Wabash Ave.\\
Terre Haute, IN 47803, USA\\
\{wollowski, berry123, winckrc, jern, voltmer, chiu, shibberu\}@rose-hulman.edu
}
\maketitle
\begin{abstract}
We are developing a system for human-robot communication that enables people to communicate with robots in a natural way and is focused on solving problems in a shared space. Our strategy for developing this system is fundamentally data-driven: we use data from multiple input sources and train key components with various machine learning techniques. We developed a web application that is collecting data on how two humans communicate to accomplish a task, as well as a mobile laboratory that is instrumented to collect data on how two humans communicate to accomplish a task in a physically shared space. The data from these systems will be used to train and fine-tune the second stage of our system, in which the robot will be simulated through software. A physical robot will be used in the final stage of our project. We describe these instruments, a test-suite and performance metrics designed to evaluate and automate the data gathering process as well as evaluate an initial data set.
\end{abstract}

\section{Introduction}
We are developing a system that learns human-robot communication in a way that is natural for humans and does not require any training of
the human user. 

Our system will initially employ gesture and speech recognition to determine a human's actions 
and intent and use them to determine appropriate robot actions in response to human action. It will eventually be extended to
include eye tracking, facial recognition and memory of past solutions to problems.

Our strategy for developing this system is fundamentally data-driven: we will use
data from multiple input sources and we will train key components 
with various machine learning techniques. 

To support our data-driven approach, we developed and will continue to develop systems 
designed to gather large data sets that are used to train various components of our system. 
These systems are both physical and virtual so as to obtain large and rich data sets.   

To automate the training as much as possible, we developed a test-suite of tasks with associated
metrics. 
 
This is an interdisciplinary project involving colleagues from Electrical and
Computer Engineering (Carlotta Berry and David Voltmer,) Psychology (Alan Jern,) Mathematics (Yosi Shibberu,)
Mechanical Engineering (Ryder Winck), Biomedical Engineering (Alan Chiu) and Computer Science (Michael Wollowski.) We are now in the second year of a 
multi-year project that represents our entry in the IBM Watson AI XPrize competition. The aim of our project is to make it possible for novice users of a robot to complete a
series of alphabet block assembly tasks under controlled conditions. By keeping humans in the loop through 
collaboration, we can realize the benefits of
machine precision and endurance in addition to human creativity and flexibility.

For training and development purposes, we chose the blocks world as our domain. In particular, we 
use wooden alphabet blocks. This domain provides a sufficiently realistic environment but 
simplifies the robotic aspects of the project, such as motion planning, grasping, and 
manipulation. It enables us to focus our efforts on the challenges involved in 
human-robot collaboration.
 
The tasks require varying degrees and types of human-robot interactions. Since our focus in
developing this project is human-robot interaction rather than robotics, we have attempted to make the tasks
and performance metrics independent of the performance characteristics of the particular
robot used. In order to gather large amounts of training data we developed a test-suite of tasks of various complexity
of interaction. The use of alphabet wooden blocks makes our test suite easy and inexpensive for others to replicate.
 
\subsection{Related Work}
Research on human-robot interaction (HRI) has long focused on both language and gesture \cite{LIU2017,Liu2017a,Mavridis2015}. 
This research has looked at table-top manipulation of objects, often blocks \cite{Matuszek2014,Bisk2016,Scalise2016,Penkov2017,Whitney2017,Lemaignan2017}. Until very recently, much of this research has focused on grounding references and spatial attributes through formulaic approaches using a limited vocabulary and a limited set of gestures \cite{Bisk2016}. Recently there has been an increased interest in data-driven approaches based on large data sets \cite{Bisk2016,Orkin2011,Ishiguro2016,Scalise2016,Mavridis2015}. Most of this work has focused on data collection through simulation and online games \cite{Bisk2016,Orkin2011,Scalise2016}. In particular the work by Bisk et al. used annotated sequences of actions of a simulated block game to train neural networks to identify the commands to move a block from one location to another \cite{Bisk2016}. Misra et al. expanded on this work by using reinforcement learning to take the language and image data from \cite{Bisk2016} and directly plan actions \cite{Misra2017}.

The work by \cite{Ishiguro2016} uses sensor data from real world human-human interaction to train a robot for HRI. In our research we hope to initially use both simulation and real world sensor data from interaction and combine this data and then expand to data gained from interaction between our algorithm and a human in both simulated and real world environments. Rather than using human description of fixed actions and scenarios \cite{Bisk2016,Scalise2016} our initial data contains interactions between two humans. However, unlike \cite{Ishiguro2016} we are using a human-as-a-robot approach which limits the abilities of the person in the "robot" role, to make their role analogous to the physical robot of our eventual system.

\section{Data}
 
The rate of progress in AI research is increasing. Much of this research is publicly available.
We believe this research coupled with exponentially decreasing hardware costs has created unique
opportunities to \emph{engineer} rather than \emph{invent} solutions. We believe that understanding human actions and intent in a spatial context is far 
too unstructured for direct programming of a solution. We wish to achieve human-robot collaboration by training components of our system with data, 
i.e. we wish for it to learn through experience.

Our domain consists of wooden blocks. All blocks are of the same size and shape and the domain is finite.
The blocks are labeled on all sides by different symbols. For some of the tasks the labels will be employed
and for some they are irrelevant. The space on a table is sufficiently large that all blocks fit on it. As
the testing suite details, we are primarily interested in human-robot collaboration and not in planning. We
should point out that there will be some amount of planning necessary on part of the robot; however, due to the
nature of our project, the planning will be quite limited.
 
\subsection{Prototype}

To help us understand and refine the problem we wish to solve, we developed an MVP prototype.  We tested the prototype with a novice user. The results were expectedly limited in scope and complexity, however, these results provided initial data and experiences that enabled us to develop our initial set of test-suites and performance metrics (see the section on~\nameref{testsuites}).  

\subsection{Transfer learning} 
\begin{figure}[H]
	\begin{center}
        \includegraphics[width=3.3in]{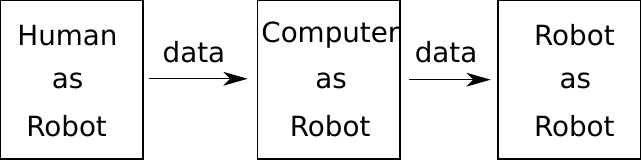}
	\end{center}
	\caption{Transfer learning data}
    \label{fig:TransferLearning}
\end{figure}

To bootstrap our data-driven system, we use transfer learning between three consecutive stages (see Figure~\ref{fig:TransferLearning}). In the first stage, a human assumes the role of the robot. In the second stage, a computer simulation is used in place of the robot. The final stage involves a robot in the physical world.

The first stage is designed to generate sufficient data to train a neural network that is then used to replace the human-as-robot. This neural network will be used to drive a computer simulation of the robot in a 2D and eventually in a 3D simulated environment. The second stage is designed to generate data that more accurately represents human-robot interaction that will occur in the final stage. In the final stage, the neural network that has been trained in the second stage is refined further using a robot.

\subsection{Human as robot} For the first stage, we developed two different systems: (i) an instrumented human who assumes the role of the robot and (ii) a web-based system in which one of the users assumes the role of the robot. 

\begin{figure}[H]
	\begin{center}
        \includegraphics[width=3.3in]{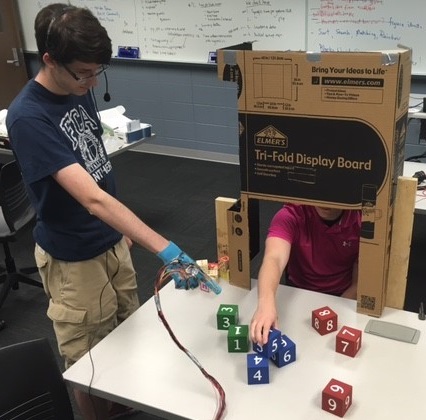}
	\end{center}
	\caption{Lab Setting of Human as Robot}
    \label{fig:HumansRobot}
\end{figure}

System (i) occurs in a laboratory and involves physical blocks. It is depicted in Figure~\ref{fig:HumansRobot}. The human assuming the role of the robot operates behind a barrier with only hands and arms visible. This is to eliminate non-verbal cues and focus the input on speech and gestures as much as possible.  Data is collected by a range of sensors: a Microsoft Kinect is used to track the human's arm gestures; a 2D camera is used to track human's face and eye gaze. The human also wears a glove with flex sensors to measure finger motion and an accelerometer is located on the top of the hand. With this system, we aim to record a rich data set of gesture and language in a 3D environment that is realistically noisy.

\begin{figure}[H]
	\begin{center}
        \includegraphics[width=3.3in]{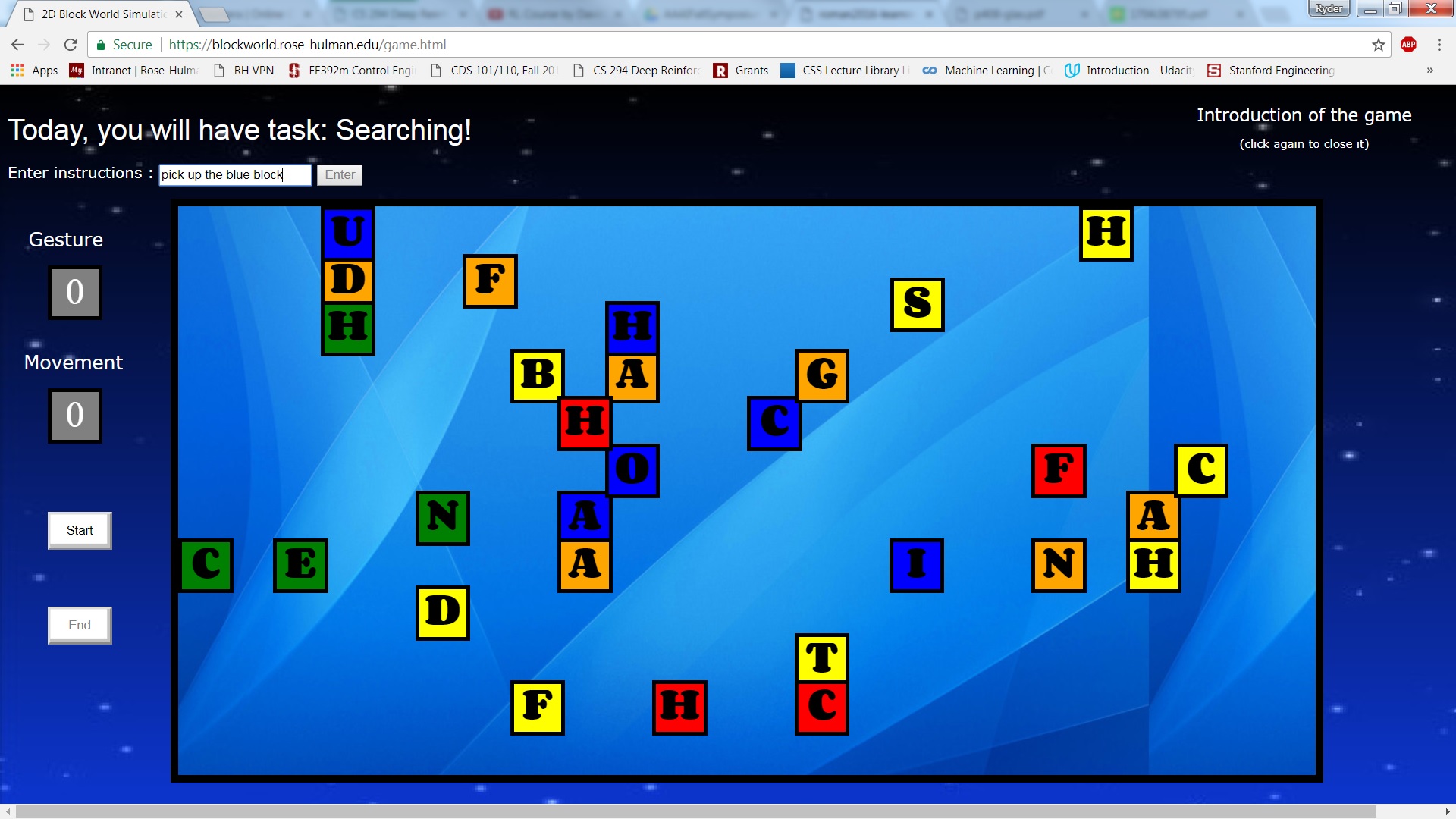}
	\end{center}
	\caption{Web-based Interaction of Human as Robot}
    \label{fig:HumansSimulation}
\end{figure}

System (ii) involves two humans also, however, they will engage in a web-based system that operates a virtual environment. The web-based interface is shown in Figure~\ref{fig:HumansSimulation}. 
It is a 2D web-based simulation of the alphabet block task. When visiting the simulation, users are automatically paired up. One user is assigned to the \emph{human role}. The human user is responsible for guiding and instructing the other user toward completing the common goal, such as sorting the blocks in a particular way. The human user may issue commands and can click on the screen as a form of gesture. Accordingly, the human user is allowed to communicate with the other user. The other user is assigned to the \emph{robot role}. The robot completes actions as instructed by the human user. Only the human-as-a-robot is able to move blocks and flip them over. The robot user is not told what the goal is and is not allowed to communicate with the human user. This arrangement is meant to mimic a real-life human-robot interaction in which the human sets the goals and directs the robot on what to do. Initial data collection has begun. This data will provide us with information about what kinds of instructions the human user provides to the other user, when in the interaction those instructions are provided, and where and how often the human user will use gestures, like pointing (simulated by mouse clicks online). 

\section{Test Suite} 
\label{testsuites}

In order to develop objective measures of progress we developed a comprehensive test suite. The tasks in the test suite below have not been attempted nor completed by our system, but are representative of the challenging tasks we expect our final system to be capable of solving. Our current test-suite contains a single construction task. Users are asked to supervise a robot to construct a prescribed structure under a time constraint. 

\begin{figure}[H]
	\begin{center}
        \includegraphics[width=3.3in]{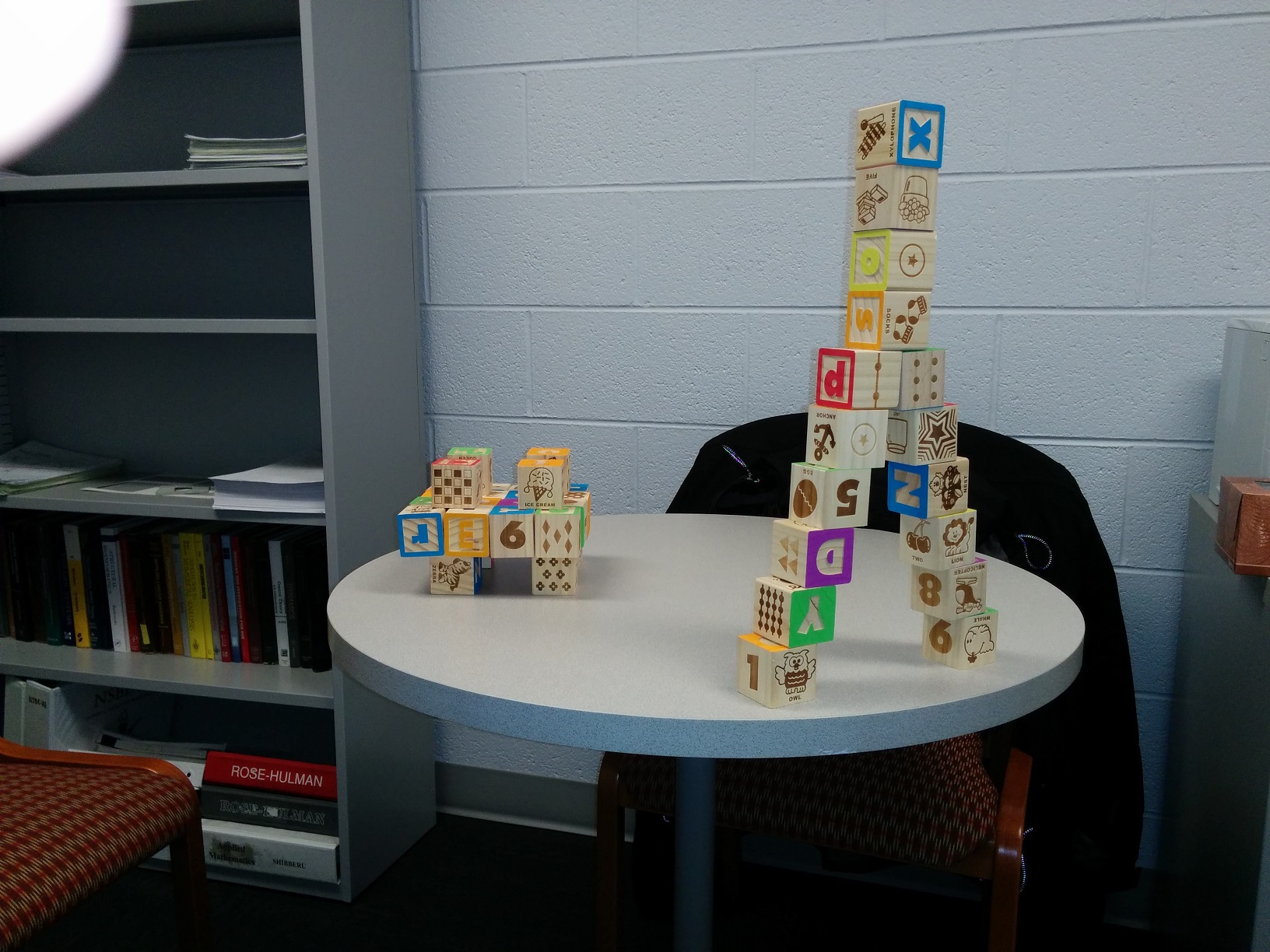}
	\end{center}
	\caption{Example of an Assembled Block Structure}
    \label{fig:TrickyToBuild}
\end{figure}

The test suite below represents our attempt to formalize what it means to have seamless 
interaction and communication with a robot. The tasks in our test suite require varying degrees and types of human-robot interactions and are intended to replicate real world tasks. One of the more challenging tasks is depicted in Figure~\ref{fig:TrickyToBuild}. Since our focus in developing this test suite is natural communication rather than robotics, we have attempted to make the tasks and performance metrics independent of the performance characteristics of the
particular robot used.
 
\subsubsection{Objectives} Each task will have one of three objectives. The human and robot must perform the following categories of tasks:
\begin{itemize}
	\item Search/sort blocks and use or store the blocks. 
	\item Replicate an existing assembled structure.
	\item Optimize some performance objective, e.g. the number of stacked blocks on a platform or the height of an assembled structure.
\end{itemize}

\subsubsection{Initial block configurations} There are three different initial block configurations that may be used. Each adds an extra layer of communication complexity to a given task. Blocks may be:
\begin{itemize}
	\item Sorted and arranged in an easily accessed one layer grid. Block operations involve retrieval of blocks from the warehouse.
	\item Scattered randomly in one layer. Block operations involve scavenging blocks from the junk yard.
	\item Arranged in a cube. Block operations involve removing blocks from the cube.
\end{itemize}

\subsubsection{Human-robot interaction} Human-robot interaction is required and enforced in each task using one of four different scenarios:
\begin{itemize}
	\item All block operations must be completed by the robot under the supervision of the human. The human may not touch the blocks.
	\item The human and robot alternate block operations.
	\item The human and robot complete block operations in an order specified by a random process.
	\item The human and robot work simultaneously, with the human using only the dominant hand. The assigned task would normally require the use of two hands. The human may use their other hand for gestures, but cannot touch blocks with it.
\end{itemize}

\subsubsection{Termination} Each task is given a certain amount of time. It terminates, when the time is elapsed.

\section{Performance Metrics} 
\label{metrics}

Issues involved in measuring the performance of human-robot interactions are discussed by several researchers~\cite{Steinfeld2006,Crandall2007,Weiss2009,Singer2011}. According to~\cite{Olsen2003}, reducing interaction effort without reducing task effectiveness is the key problem in improving the performance of interfaces that mediate human-robot interactions. 

During the competition, we improved our originally proposed metric so that it is both more accurate and easier to implement across our software and hardware projects.  Our new metric defines human-robot communication efficiency to equal: 

\[
\text{communication efficiency} = \frac{\text{task efficiency}}{\text{communication effort}}
\]
where
\[
\text{task efficiency} = \frac{\text{useful work output}}{\text{work input}}
\]
and communication effort is a linear combination of natural language communication effort and gesturing effort of the human user. 

In order to quantify task efficiency, we first define construction error to equal the absolute coordinate error ($L^1$ metric) between the human user's construction and the desired target construction. This error is computed after the centroids of the
human user and the target construction have been aligned. In
other words, pure translation of the human user’s construction
does not reduce construction accuracy. Task efficiency is then defined to equal the reduction in absolute coordinate error achieved by the human user divided by the total amount of block movement during the time allotted for the task. 
\[
\text{task efficiency}=\frac{\text{reduction in coordinate error}}{\text{total block movement}}
\]

Natural language effort is defined to equal a normalized word count of the number of words spoken by the human user during the time allotted for the task. For simulated environments, gesture effort is defined to equal a normalized count of human user mouse clicks. For our physical laboratory setting, accelerometer measurements are used to estimate the total human user's arm movements during the time allotted for the task. The arm movement total is normalized by the human user's arm length. 

The human user is given feedback in the form of a task completion score. The goal of the human user is to maximize the task completion score in the time allotted for the task. The task completion score is defined to equal
\[
\text{completion score} =\frac{\text{initial error}-\text{current error}}{\text{initial error}}.
\]

\section{Test Suite Automation}

In order to train key software of our system, it is imperative that we automate the training as much as possible. We will use the performance metrics towards this end. In order to evaluate the complexity of a task, we need to assess and measure block operations as well as define similarity between desired task outcome and the actual outcome. In order to automate test suite computations and to evaluate the quality of the human-robot interaction, we need to evaluate the gesture and voice input. To this end, we plan to use 3D cameras, user worn accelerometers and data generated by the robot sensors. 

Quantifying human gesture effort is a challenging task. We will use machine learning methods to segment and classify the human gestures we observe during task completion once we have collected a sufficiently large data set. Assessing the voice input will rely on the parsed sentence produced by the various components. We can measure the length of the dialog, the length of the sentences and to a certain degree, the complexity of a sentence.

\section{Evaluation of Preliminary Data}

We are currently gathering data from the web version of the human-as-robot games. We have collected data from over 200 games so far with a total of over 7,000 human user and human-as-robot actions.

Recall from the section on~\nameref{metrics} that our objective is to optimize human-robot communication efficiency.  In this context, we 
defined task efficiency as the reduction
in the absolute value of coordinate error between initial block configuration and the specified target block
configuration. More formally, we define the error of the initial, given block configuration as:
\[
\text{error\_init} = \sum_{k=0}^{num blocks} |x_{k}^{init} - x_{k}^{target}| + |y_{k}^{init} - y_{k}^{target}|
\]
 
Error is computed after centroids are aligned so that the error is translation invariant.
The final error, {\em error\_final} is calculated after the game is over. We can now define the completion score more formally as:
\[
\text{completion score} =\frac{\text{error\_init}-\text{error\_final}}{\text{error\_init}}
\]

Note that the completion score will be negative if the final configuration error is
larger than the initial configuration error. 

We define the {\em dist\_moved} is the sum of the L1 or cityblock distances the blocks have been moved. We can now define task efficiency as follows:
\[
\text{task efficiency}  =\frac{\text{error\_init}-\text{error\_final}}{\text{dist\_moved}}
\]

Let {\em eff\_word} and {\em eff\_gesture} be the number of words and number of gestures divided by their respective averages to
normalize them. We can now define communication effort as:
\[
\text{communication effort} = \text{0.8 eff\_word} + \text{0.2 eff\_gesture} 
\]

\subsection{Filtering Games}
Low communication effort suggests either a lack of serious effort or a hidden communication channel
between the human user and the human-as-robot. For example, a single user can play both roles on
their computer screen by opening up two browsers. Communication of information in this case is not
required as one player is playing both roles. Another hidden communication channel occurs when
both players are in the same room and are able to identify and each other and communicate
physically. We removed games with communication effort less than 0.5.

\begin{figure}[H]
	\begin{center}
        \includegraphics[width=3.3in]{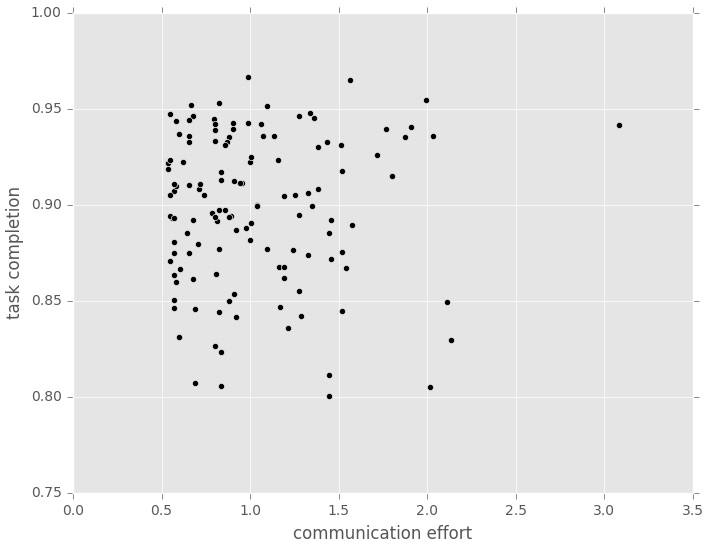}
	\end{center}
	\caption{Effort and completion after filtering out games with invalid data}
    \label{fig:filtered_games}
\end{figure}

Low completion rate suggests a lack of interest in the game or players just exploring game commands
and options. We removed games with completion rate less than 0.8. This resulted in 127 games. 
The plot of those games is shown in figure~\ref{fig:filtered_games}.

\subsection{Words vs Gestures}

\begin{figure}
	\begin{center}
        \includegraphics[width=3.3in]{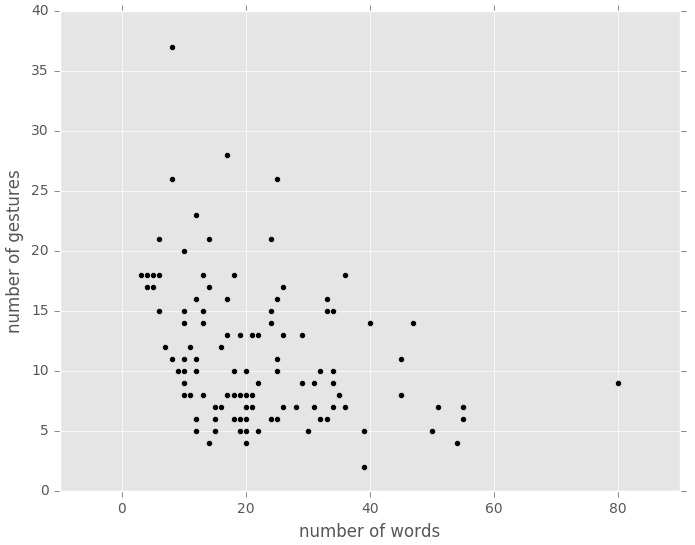}
	\end{center}
	\caption{Relationship between number of words and number of gestures used}
    \label{fig:wordsVgestures}
\end{figure}

Figure~\ref{fig:wordsVgestures} suggests that there is an inverse relationship between the number of words used
versus the number of gestures used.

\subsection{Task Completion vs Communication}

There does not appear to be any correlation between the task completion metric and the
communication performance metric. This results is desirable, because we do not want the user to focus
on learning how to improve our performance metric. We want our performance metric to measure
natural (unlearned) communication effectiveness. Asking the user to focus on the completion metric likely 
helps avoid users "gaming" our performance metric. In our own experience, the performance in playing improved with experience. 
We suspect that this holds for all users, but at this point to not have the data to support this claim. Our IRB is written in a way 
that does not permit us to store identifiable information and thus, we cannot identify repeat users.

\begin{figure}[H]
	\begin{center}
        \includegraphics[width=3.3in]{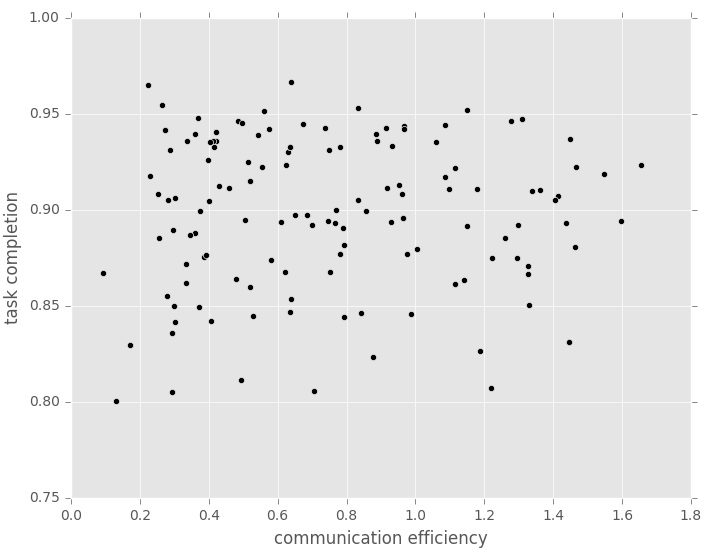}
	\end{center}
	\caption{Relationship between task completion and communication}
    \label{fig:taskCompletionVCommunication}
\end{figure}

\subsection{Gestures vs. Words}

\begin{figure}
	\begin{center}
        \includegraphics[width=3.3in]{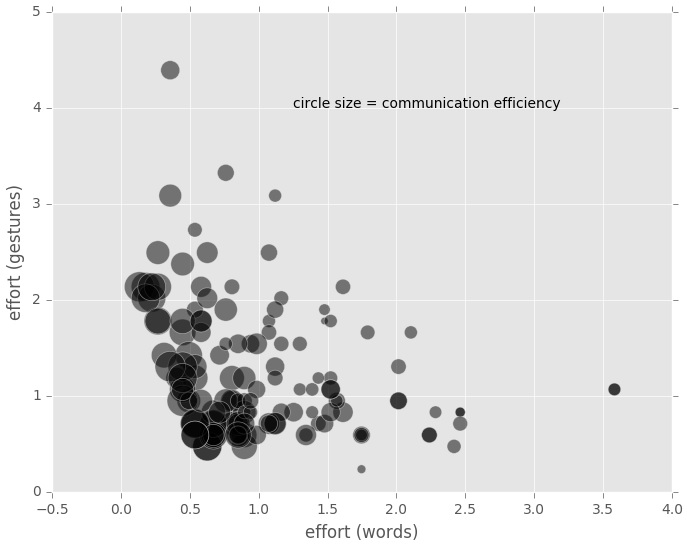}
	\end{center}
	\caption{Gestures should be favored over words}
    \label{fig:favorGestures}
\end{figure}

Figure~\ref{fig:favorGestures} suggests better performance can be achieved by using more gestures than words. The size of a circle 
represents performance. The graph is dependent on the relative weights given to words and gestures in the effort metric. The suggestion 
makes sense because the score of the game is dependent on the accuracy of placing blocks at the desired location.

\section{Conclusions}

We described our data-driven approach towards engineering a system that is concerned with seamless human-robot communication. We explained the use of transfer learning to boot-strap our system with increasingly realistic data. We detailed two initial data gathering systems. Both systems involve two human beings; one of them assuming the role of a robot. One of the systems takes place in a lab setting and is designed to gather more noisy and realistic initial data. In it, the person assuming the role of the robot is instrumented. The other system is set in a virtual environment and is designed to gather data with which to train the second stage of our system. In the second stage, the robot will be simulated through software rather than a person. The second stage is designed to gather more realistic data. Only in the final stage will 
we use and train a physical robot. We described a test suite that is designed to test and train our system and in a sense defines complexity for human-robot interaction. We presented performance metrics as well as an evaluation of our initial data set. From the evaluation, we learned that our data gathering instruments are sound: there does not appear to be any correlation between the task completion metric and the communication performance metric. We observed that there is an inverse relationship between the number of words used versus the number of gestures used. Based on our evaluation metric, better performance can be achieved by using more gestures than words.

The immediate future is about generating yet more data through our human-as-robot systems. We aim to gather data from 1,000 games, primarily from the virtual system. This will enable us to proceed to the second stage in which a software bot will assume the role of the robot. For this stage, we will set-up and train the bot with the data gathered during the first stage. The bot will be deployed in the 2D simulation software. In order to be prepared for the third stage of the project, one in which a physical robot will be used, we will additionally develop and train a 3D software simulation of our system. This latter simulation will conduct automated performance measure computations. 
 
\section*{Acknowledgements}
We would like to acknowledge the contributions of the following students who worked on developing the MVP of our human-robot collaboration platform: Devon Gray Adair, Chris Collinsworth, Zach Dougherty, Corie Lynn Ewoldt, Jackson Fairfield, Bryan D Gish, Manoj Kurapati, Lecea Sun and Zhihao Xue. We  would like to acknowledge the contributions of the following students who developed the human-as-robot data gathering systems: Qikai Huang, Ryan DeCramer, Yunyan Ding, Avery Pratt, Aaradhana Bharill and Shay Merley.
 
\bibliography{ms}
\bibliographystyle{aaai}
\end{document}